\documentclass[10pt,conference]{IEEEtran}
\usepackage{epsfig,makeidx,color}
\usepackage{amsmath,amssymb,amsthm,bbm}
\usepackage{cite,graphicx}
\usepackage{enumerate}

\def\cK{{\cal K}}

\def\cG{{\cal G}}

\def\rT{{\rm T}}

\def\uE{{\mathbb E}}

\newtheorem{mylemma}{\bf Lemma} % [section]
\def\be{ \begin{equation} }
\def\ee{ \end{equation} }
\def\bea{ \begin{eqnarray} }
\def\eea{ \end{eqnarray} }

\def\bs{{\bf s}}

\def\bB{{\bf B}}

\def\b0{{\bf 0}}

\def\cS{{\cal S}}

\ifCLASSOPTIONonecolumn
  \newcommand{\figwidth}{0.60\columnwidth}
\else
  \newcommand{\figwidth}{0.90\columnwidth}
\fi

\begin{document}

\title{A Game-Theoretic Approach for NOMA-ALOHA}

\author{Jinho Choi \\
School of Electrical Engineering and Computer Science \\
Gwangju Institute of Science and Technology (GIST) \\
\emph{Email: jchoi0114@gist.ac.kr}}

%\thanks{
%School of Information and Communications,
%Gwangju Institute of Science and Technology (GIST),
%Gwangju, 500-712, Korea.}}
%This research was supported by GIST Research Fund 
%(for the Project of ``Living Energy") in 2016.}}

\date{today}

\maketitle

\begin{abstract}
Non-orthogonal multiple access (NOMA) 
can improve the spectral efficiency 
by exploiting the power domain 
and successive interference cancellation (SIC),
and it can be applied to various transmission schemes
including random access that plays a crucial role
in the Internet of Things (IoT) to support connectivity
for a number of devices with sparse activity.  In this paper, we formulate 
a game when NOMA is applied to ALOHA 
to decide the transmission probability.
We consider a payoff function based on an energy-efficiency
metric and drive the mixed strategy Nash equilibrium (NE).
\end{abstract}

{\IEEEkeywords
random access, non-orthogonal multiple access, game theory}

\section{Introduction} \label{S:Intro}

Non-orthogonal multiple access (NOMA) 
is to exploit the power domain and allows
multiple users to share the same radio resource.
As in \cite{Choi08G}, successive interference cancellation (SIC)
can be employed in NOMA to effectively decode signals in the
presence of interference.
Since NOMA can 
improve the spectral efficiency of multiuser systems,
it has been widely investigated to 
and considered for 5th generation (5G) systems
\cite{Dai15, Higuchi15, Ding_CM, Choi17_ISWCS}.

For downlink NOMA,
superposition coding and SIC 
are employed,
where the power allocation becomes crucial to guarantee
successful decoding with SIC.
Beamforming with user clustering
is considered in \cite{Kim13}, the sum rate
optimization is investigated with a minorization-maximization method
in \cite{Hanif16},
and a generalized NOMA beamforming
approach is studied in \cite{Choi17_JCN}
in order to take into account the spatial correlation.

NOMA can also be employed for uplink transmissions.
For uplink NOMA, the power allocation to 
guarantee successful SIC is studied in \cite{Imari14}.
In \cite{ChoiJSAC, Choi17_L}, 
NOMA is applied to ALOHA, which is a random access
scheme \cite{BertsekasBook},
with different power levels. With NOMA,
since more (virtual) channels can be available,
the throughput of ALOHA can be improved, which implies
that NOMA-ALOHA can support more users or devices
in the Internet of Things (IoT)
with a limited bandwidth.

In general, random access schemes 
are suitable for a number of users or nodes with sparse active
as signaling overhead to allocate radio resources
is not required. Thus, random access is considered
for machine-type communications (MTC) in \cite{Shar15}, and 
employed for standards
as in \cite{3GPP_MTC, 3GPP_NBIoT}.
While signaling overhead is low in random access,
each node or user has to decide its transmission parameters.
In \cite{Menache08, CLD10, Cohen16},
it is shown that the notion of game theory \cite{Maschler13}
becomes useful to locally optimize transmission
parameters for random access schemes,
since random access can be seen as a noncooperative
game.

In this paper, we propose a game-theoretic
approach for NOMA-ALOHA proposed in 
\cite{ChoiJSAC} to decide the transmission
probability or the mixed strategy for transmissions.
In particular, we formulate a NOMA-ALOHA game based
on an energy-efficiency metric for the payoff function and derive
the mixed strategy Nash equilibrium (NE)
for each user's transmission or access probability.

\subsubsection*{Notation}
Matrices and vectors are denoted by upper- and lower-case
boldface letters, respectively.
The superscript $\rT$ denotes the transpose.
$\uE[\cdot]$ denotes the statistical expectation.
%$\cC \cN(\ba, \bR)$
%represents the distribution of
%a circularly symmetric complex Gaussian 
%(CSCG) random vector with mean vector $\ba$ and
%covariance matrix $\bR$.

\section{Single-Channel Two-Person Game}	\label{S:2P}

In this section, we consider 
NOMA-ALOHA with single channel
in order to demonstrate
how a game-theoretic approach can be employed to decide
transmission probabilities. 

\subsection{Two-Person NOMA-ALOHA Game}

Suppose that there are two users\footnote{Throughout the paper,
we assume that users and players are interchangeable.} 
for multiple access
(or uplink transmissions) with a single channel
and a receiver or base station (BS).
We assume that NOMA is employed with two different power
levels. Thus,
the number of transmission strategies for each user
is 3 as follows:
$$
\cS_k = \cS = \{H, L, 0\}, \ k = 1, 2,
$$
where the subscript $k$ is the index for users,
$H$ and $L$ represent the high and low transmit powers, respectively,
and $0$ represents no transmission.
Denote by $s_k$ the strategy of user $k$.
In addition, $\bs_{-k}$ represents the set of the strategies
of the users except user $k$,
i.e., $\bs_{-k} = \{s_1, \ldots, s_{k-1}, s_{k+1}, \ldots, s_K\}$,
if there are $K$ users. For two-person games,
$\bs_{-1} = s_2$ and $\bs_{-2}  = s_1$.

The payoff function of user $k$ is given by
\begin{align}
u_k (s_k, \bs_{-k}) & = u(s_k,\bs_{-k}) \cr
& = R (s_k, \bs_{-k}) - C(s_k), \ k = 1,2,
	\label{EQ:payoff}
\end{align}
where $R(s, s^\prime)$ is the reward function of successful transmission
and $C(s)$ is the cost 
function of transmission strategy. In particular, we consider
the following reward function:
\be
R (s, s^\prime) = \left\{
\begin{array}{ll}
W, & \mbox{if $s \ne s^\prime$ and $s \in \{H,L\}$} \cr 
0, & \mbox{o.w.,} \cr
\end{array}
\right.
	\label{EQ:RW}
\ee
where $W> 0$ is the reward of successful transmission for a user.
Due to NOMA, if the user of interest chooses $s = H$, while the other user
chooses $s^\prime = L$ or $0$, the user of interest can successfully transmit
his signal. 
%For the case of $(s,s^\prime) = (L,H)$ or $(L, 0)$,
%the user of interest has a successful transmission.
Thus, the main difference of NOMA-ALOHA from conventional ALOHA
is that the BS is able to recover the signals from two users simultaneously
as long as one user employs strategy $H$ and the other user
adopts strategy $L$ as in \eqref{EQ:RW}.
Note that in \eqref{EQ:RW},
as in conventional ALOHA, if two users choose 
$(H,H)$ or $(L,L)$, we assume collision and the BS is not able to
receive any signal \cite{BertsekasBook}.

For the cost function, we can consider the following assignment as an example:
$$
C(H) = 2, \ C(L) = 1, \ \mbox{and} \ C(0) = 0,
$$
because strategy $H$ requires a higher transmit power than
strategy $L$.

It is noteworthy that the payoff function in \eqref{EQ:payoff}
can be seen as an energy-efficiency metric, which is widely used
in wireless systems, e.g., \cite{Saraydar02} for power control game.
To see that the payoff function in \eqref{EQ:payoff} is an
energy-efficiency metric, 
we can consider the logarithm of the ratio of the spectral efficiency
or throughput
to the transmit power as follows:
$$
\ln \frac{\mbox{Throughput}}{\mbox{Transmit Power}} 
= \ln(\mbox{Throughput})- \ln(\mbox{Transmit Power}),
$$
where $\ln(\mbox{Throughput})$ becomes
the reward function and $\ln(\mbox{Transmit Power})$ becomes
the cost function in \eqref{EQ:payoff}.

The two-person NOMA-ALOHA game 
has a strategic form of triplet: 
{\it 1)} $\cK = \{1,2\}$, where $\cK$ represents
the set of users or players; {\it 2)} $\cS_k = \cS$, where $\cS_k$
denotes the set of strategies of user $k$; {\it 3)} the payoff functions 
in \eqref{EQ:payoff}.
That is, the two-person NOMA-ALOHA game is
given by $\cG = \{\cK, \{\cS_k\}_{k=1}^K, \{u_k\}_{k=1}^K\}$
with $K = 2$.
The resulting game is symmetric,
that is, both players have the same set of strategies, 
and their payoff functions satisfy $u_1(s_1,s_2) = u_2(s_2,s_1)$
for each $s_1,s_2 \in S$ \cite{Neumann44}.
Furthermore, its bimatrix can be found as in
Table~\ref{TBL:2P}.

\begin{table}[thb]
\caption{Bimatrix of two-person NOMA-ALOHA game.}
\begin{center}
\begin{tabular}{c|ccc} 
 & $H$ & $L$ & 0 \\ \hline
 $H$ & $(-2,-2)$ & $(W-2, W-1)$ & $(W-2,0)$ \\
 $L$ & $(W-1,W-2)$ & $(-1, -1)$ & $(W-1,0)$ \\
 $0$ & $(0,W-2)$ & $(0, W-1)$ & $(0,0)$ \\
\end{tabular}
\end{center}
\label{TBL:2P}
\end{table}

It is noteworthy that the two-person NOMA-ALOHA game
can be seen as a 
generalization of a multiple access game in \cite{Felegyhazi06}
with the notion of NOMA \cite{ChoiJSAC}.
The multiple access game in \cite{Felegyhazi06} has two strategies
for each user: Transmit (T) and Quite (Q). Strategy T is further
divided into $H$ and $L$ in the two-person NOMA-ALOHA game,
while strategy Q becomes $0$.

\subsection{Finding NEs}

In this subsection, we find NEs 
of the two-person NOMA-ALOHA game
and show that the NEs 
depend on the reward of transmission, $W$.

If $W \le 2$, there exist pure strategy NEs \cite{Bacci16, Maschler13},
denoted by $\{s_k^*\}$,
which are characterized by
$$
u_k (s_k^*, s_{-k}^*)\ge
u_k (s_k, s_{-k}^*) \ \mbox{for all} \ s_k \in \cS_k, \ k \in \cK.
$$
For example, for $0 \le W < 1$, 
from Table~\ref{TBL:2P}, we can see that $(s_1, s_2) = (0,0)$
is the pure strategy NE. That is, if the reward of successful
transmission, $W$, is sufficiently small (compared to the cost
of transmissions), 
the users do not want to transmit signals and non-transmission
strategy (i.e., $s_k = 0$) becomes NE.
For
$1 \le W \le 2$,
the pure strategy NEs are $(s_1, s_2) = (0, L)$ and 
$(s_1, s_2) = (L, 0)$.
If $W > 2$, there is no pure strategy NE.

In general, we are interested in 
mixed strategy NEs as randomized strategy can be well employed
for random access.
In order to find the mixed strategy NEs, 
the principle of indifference \cite{Maschler13} can be used.
Since the two-person NOMA-ALOHA game is symmetric,
it suffices to find one user's mixed strategy NE.
To this end, let
$a$ and $b$ denote the probabilities to choose $H$ and $L$, respectively.
Thus, a mixed strategy is represented by $\sigma = (a, b, 1-a- b)$, 
where $a + b \le 1$.

For convenience, let $\bB$ denote the payoff matrix for the row user
in Table~\ref{TBL:2P}. Let $[\bB]_{n,m} = B_{n,m}$.
According to the principle of indifference,
the row user has the same expected payoff for any pure strategy
when the column user employs the mixed strategy NE.
Thus,
it follows
\begin{align}
U & = a^* B_{1,1} + b^* B_{1,2} + (1-a^* - b^*) B_{1,3} \cr
& = a^* B_{2,1} + b^* B_{2,2} + (1-a^* - b^*) B_{2,3} \cr
& = a^* B_{3,1} + b^* B_{3,2} + (1-a^* - b^*) B_{3,3},
	\label{EQ:id}
\end{align}
where $U$ is the expected payoff of the row user 
and $(p^*, q^*, 1 - p^* - q^*)$ is the mixed strategy NE.
From \eqref{EQ:id}, we can have two equations for two unknown 
variables, $a^*$ and $b^*$. 
In addition, since $a^* + b^* \le 1$, we can find 
$a^*$ and $b^*$.

Noting that $B_{3,i} = 0$ for $i = 1,2,3$ (as the row user
does not transmit) from Table~\ref{TBL:2P}, 
we can see that $U = 0$ if $1 - (a^* + b^*) > 0$
(i.e., the probability of non-transmission or strategy $0$ is greater than 0).
In this case, we can have closed-form expressions
for $a^*$ and $b^*$ from \eqref{EQ:id} as follows:
\begin{align}
a^* = \frac{W-2}{W} \ \mbox{and} \ 
b^* = \frac{W-1}{W}.
	\label{EQ:ab}
\end{align}
The above solution is valid when $2 \le W < 3$ 
since $a^*, b^* \ge 0$ and $1 - (a^* + b^*) > 0$
are required. If $W = 3$, we can see that
$a^* + b^* = 1$, which means that
the probability of non-transmission is 0. That is,
strategy 0 is not used if the reward of successful
transmission, $W$, is sufficiently large. 
Thus, for $W \ge 3$,
\eqref{EQ:id} is reduced to 
\begin{align}
U & = a^* B_{1,1} + b^* B_{1,2} + (1-a^* - b^*) B_{1,3} \cr
& = a^* B_{2,1} + b^* B_{2,2} + (1-a^* - b^*) B_{2,3}.
\end{align}
Then, after some manipulations, we have
\begin{align}
a^* = \frac{W-1}{2W} \ \mbox{and} \ 
b^* = \frac{W+1}{2W}, \ W \ge 3.
	\label{EQ:abW3}
\end{align}
%This shows the asymptotic behavior
%of two-person NOMA ALOHA that
%$a^*, b^* \to \frac{1}{2}$ as $W \to \infty$.
%That is, as the reward of successful transmission, $W$,
%increases, the probabilities of $s_k = H$ and
%$s_k = L$ approach $\frac{1}{2}$.

We now consider the case that $W < 2$. If $W < 2$,
the reward of successful transmission is so small that 
high-power transmission is not desirable. Thus, 
$a^* = 0$ (which is the case that $W = 2$ as shown in \eqref{EQ:ab}).
Thus,
\eqref{EQ:id} is reduced to 
\begin{align}
U = b^* B_{2,2} + (1 - b^*) B_{2,3} = 0,
\end{align}
which leads to
\be
b^* = \left\{
\begin{array}{ll}
\frac{W-1}{W}, & 1 \le W \le 2 \cr
0, & 0 \le W < 1. \cr
\end{array}
\right.
\ee

\begin{figure}[thb]
\begin{center}
\includegraphics[width=\figwidth]{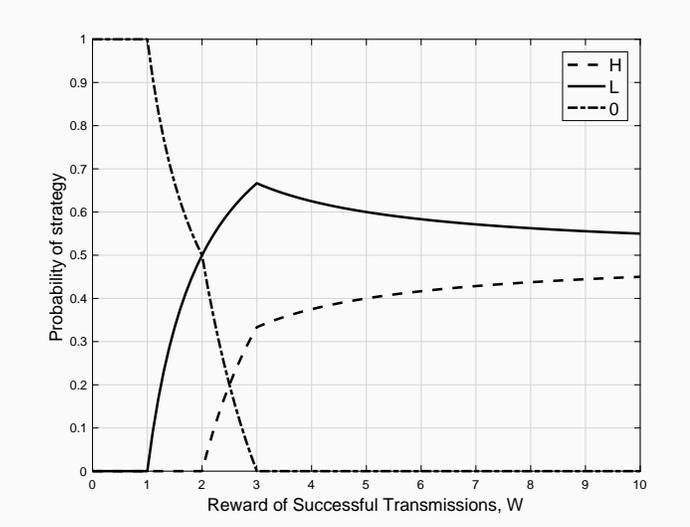}
\end{center}
\caption{The mixed strategy NE, $\sigma^* = 
(a^*, b^*, 1- a^* - b^*)$, for different values
of the reward of successful transmission, $W$.}
        \label{Fig:eplt1}
\end{figure}

In Fig.~\ref{Fig:eplt1}, we show the mixed 
strategy NE, $\sigma^* = 
(a^*, b^*, 1- a^* - b^*)$, for different values
of the reward of successful transmission, $W$.
As shown in 
Fig.~\ref{Fig:eplt1}, we can see that
the probability of strategy $L$ is higher than 
the probability of strategy $H$ as 
strategy $H$ has a higher cost than strategy $L$.
In addition, as $W$ increases, 
the probability of strategy $0$ decreases.
That is, as the reward of successful transmission
increases, the users tend to transmit signals.
Note that as $W \to \infty$, $a^* = b^* \to \frac{1}{2}$
from \eqref{EQ:abW3}, i.e.,
the users always transmit.

When users always transmit in conventional ALOHA, 
there are collisions with probability
(w.p.) 1 and the throughput 
(i.e., the average number of successfully
transmitted packets)  becomes 0. However, in NOMA-ALOHA,
the throughput 
does not approach $0$ although collisions happen
thanks to NOMA.  In the two-person NOMA-ALOHA,
as $W\to \infty$, the asymptotic throughput approaches $1$, because
the BS is able to recover 
the two users' signals simultaneously as long as $(s_1, s_2)
= (H, L)$ or $(L, H)$, i.e., 
\begin{align*}
{\rm Throughput} & = 
2 \times \Pr \left((s_1, s_2) = (H, L) \ \mbox{or} \
(L, H) \right) \cr
& = 2 \times \frac{1}{2} = 1, \ W \to \infty.
\end{align*}

\subsection{Average Payoff Maximization}

In this subsection, we consider a different approach
that is not based on noncooperative game.

\begin{figure}[thb]
\begin{center}
\includegraphics[width=\figwidth]{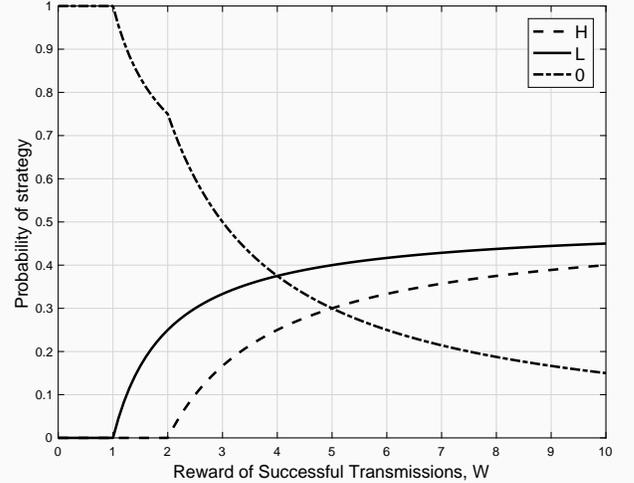}
\end{center}
\caption{The mixed strategy that maximizes the average payoff
for different values
of the reward of successful transmission, $W$.}
        \label{Fig:eplt2}
\end{figure}

Suppose that a mixed strategy
is used and the two users have the same mixed strategy,
$\sigma = (a, b, 1-a - b)$, and use it independently
(as no cooperation is assumed).
From \eqref{EQ:payoff}, for a given mixed strategy, the average
payoff is given by
\begin{align}
\bar u(a,b) & =  \uE[u(s_1, s_2)] \cr
& = \uE[R(s_1, s_2)] - \uE[C(s_1)] \cr
& = W \left(a (1-a) + b(1-b) \right) - 2 a - b.
	\label{EQ:bu}
\end{align}
Since $\bar u(a,b)$ is concave in $a$ and $b$,
the maximization of the average payoff can be carried out.
In Fig.~\ref{Fig:eplt2},
we show the optimal mixed strategy, denoted by 
$\hat \sigma= (\hat a,\hat b,1-\hat a- \hat b)$
that maximizes the average payoff
for different values
of the reward of successful transmission, $W$.
We can see that 
the optimal mixed strategy that maximizes the average payoff
is similar to  the mixed strategy NEs in Fig.~\ref{Fig:eplt1},
although both the mixed strategies are not the same for $W \ge 1$.
For example, for $0 \le W\le 1$, the probability of strategy $0$
is 1 in both the mixed strategies. In addition,
we see that the probability of strategy $L$ is higher
than or equal to that of strategy $ H$,
while the two probabilities approaches $\frac{1}{2}$ as $W \to \infty$
in  both the mixed strategies. 

\begin{figure}[thb]
\begin{center}
\includegraphics[width=\figwidth]{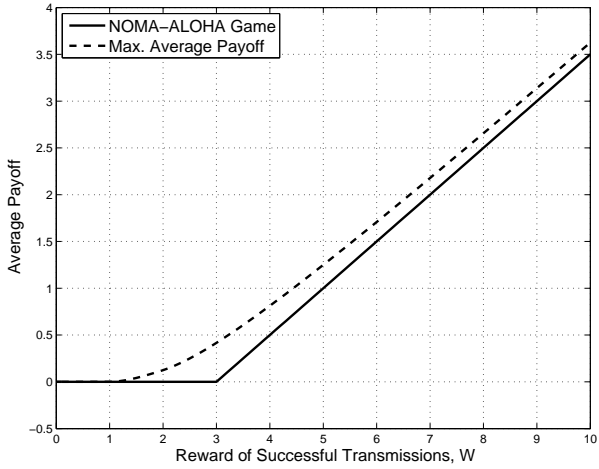}
\end{center}
\caption{The maximum average payoff and the average payoff
of the mixed strategy NE of the two-person NOMA-ALOHA for
different values of $W$.}
        \label{Fig:apoff}
\end{figure}

Fig.~\ref{Fig:apoff} shows the average payoff functions
of the two mixed strategies, $\hat \sigma$ and $\sigma^*$,
for different values of $W$. Clearly, 
$\hat \sigma$ provides the highest 
average payoff and higher than that of $\sigma^*$.
However, since $\hat \sigma$ is not a mixed strategy NE,
any user who uses a slightly different mixed strategy from 
$\hat \sigma$ can have a higher average payoff at the cost
of the degraded average payoff of the other user who uses $\hat \sigma$.

The ratio of the payoff with $\hat \sigma$
to that with  $\sigma^*$ can be seen as a price of
anarchy (PoA) \cite{Koutsoupias99, Nisan07}.
If both the users trust each other, they
can employ $\hat \sigma$. On the other hand, if there is no trust, each
user may need to employ $\sigma^*$, which is NE,
and has a worse average payoff 
than that can be obtained with $\hat \sigma$,
i.e., the PoA is less than 1.
However, as $W \to \infty$, from \eqref{EQ:bu},
we can see that $\hat a$ and $\hat b$
become $\frac{1}{2}$, which is the same as the asymptotic
mixed strategy NE, $(a^*, b^*)$, with $W \to \infty$,
and the PoA approaches 1.

\section{Multi-User NOMA-ALOHA Game}

In this section, we generalize the two-person NOMA-ALOHA game
that was introduced in Section~\ref{S:2P}
with more users.

Suppose that there are $K \ge 2$ users.
Since each user has the same payoff function,
the NOMA-ALOHA game is symmetric and 
the principle of indifference \cite{Maschler13} can be employed
in order to find the mixed strategy NE.

For convenience, let the user of interest is user $k$.
Denote by $U(H)$, $U(L)$, and $U(0)$ the payoff values of user 1 if 
user 1 chooses $s_k = H$, $L$, and $0$, respectively, 
when the other users have the same mixed strategy,
$(a, b, 1 - a - b)$.
Let $q_k (s)$
be the probability that the other users 
do not employ strategy $s$.
Since each user chooses 
a strategy independently, we have
\begin{align}
q_k (H) & =  \prod_{i \ne k} \left(1 - a \right) 
= \left(1 - a \right)^{K-1} \cr
q_k (L) & =  \prod_{i \ne k} \left(1 - b \right)
= \left(1 - b \right)^{K-1}.
\end{align}
From this,
we show that
\begin{align}
U(H) & = -C(H) \left( 1 - q_k (H) \right) + (W-C(H)) q_k (H) \cr
U(L) & = (W-C(L)) q_k (L)- C(L) \left( 1 - q_k (L) \right),
	\label{EQ:UHL}
\end{align}
while the payoff of user $k$ becomes $U(0) = 0$ when $s_k = 0$.

Like the analysis in Section~\ref{S:2P},
we can see that
if $W < 1$, both the probabilities of strategies $H$ and $L$
are to be 0 for NE. Thus, $(a^*, b^*, 1- a^*-b^*) = (0,0,1)$.
For $1 \le W < 2$,
strategy $H$ cannot be applied for the mixed
strategy NE (i.e., $a^* = 0$). 
In this case, since we need to have
$$
U(L) = U(0),
$$
the resulting mixed strategy NE becomes
$$
(a^*, b^*, 1- a^* - b^*) = \left(0, 1 - 
\left( \frac{1}{W} \right)^{ \frac{1}{K-1} }, 
\left( \frac{1}{W} \right)^{ \frac{1}{K-1} } \right). 
$$
The mixed strategy NE for $W \ge 2$ can be shown as follows.

\begin{mylemma}
Let 
\be
W^* = \left(1 + 2^{\frac{1}{K-1}}
\right)^{K-1}.
	\label{EQ:Ws}
\ee
For $2 \le W < W^*$, 
we have
\be
(a^*, b^*) = \left(1 - 
\left(1- \frac{2}{W} \right)^{ \frac{1}{K-1} }, 
1-\left(\frac{1}{W} \right)^{ \frac{1}{K-1} }\right).
	\label{EQ:ab2WW}
\ee
For $W \ge W^*$,
$b^*$ is the solution of
\be
W b^{K-1} = W (1 - b)^{K-1} +1,
	\label{EQ:Wb}
\ee
while $a^* = 1- b^*$.
\end{mylemma}
\begin{IEEEproof}
Suppose that 
the probability that a user employs strategy 0 is not zero,
while $a, b > 0$. Then, by the principle of indifference,
we need to have
$U(H) = U(L) = U(0) = 0$.
From \eqref{EQ:UHL}, 
it can be shown that
$$
W q_k (H) -2  = W q_k (L) - 1 = 0.
$$
Thus, we have
\begin{align}
a^* = 1 - \left(\frac{2}{W} \right)^{\frac{1}{K-1}} \ \mbox{and} \
b^* = 1 - \left(\frac{1}{W} \right)^{\frac{1}{K-1}},
\end{align}
which is given in \eqref{EQ:ab2WW}.

However, 
if $W$ is sufficiently large (i.e., for a sufficiently
large reward of successful transmission),
the probability that a user employs strategy 0 becomes zero
or $a^* + b^* = 1$.
The corresponding $W$
is the solution of $1 = a^* + b^*$ or
\be
1 = 
\left(\frac{2}{W} \right)^{ \frac{1}{K-1} } +
\left(\frac{1}{W} \right)^{ \frac{1}{K-1} },
\ee
which is $W^*$ in \eqref{EQ:Ws}.
In this case (i.e., $W \ge W^*$),
we only need to have $U(H) = U(L)$
with $a^* + b^* = 1$.
Thus, we have
$$
W q_k (H) - 2 = W q_k (L) -1  \ \mbox{and} \ a^* + b^* = 1,
$$
which can also be expressed as
\eqref{EQ:Wb} in terms of $b$ only.
Clearly, the solution of \eqref{EQ:Wb} is $b^*$,
and $a^*$ becomes $1 - b^*$,
which completes the proof.
\end{IEEEproof}

\begin{figure}[thb]
\begin{center}
\includegraphics[width=\figwidth]{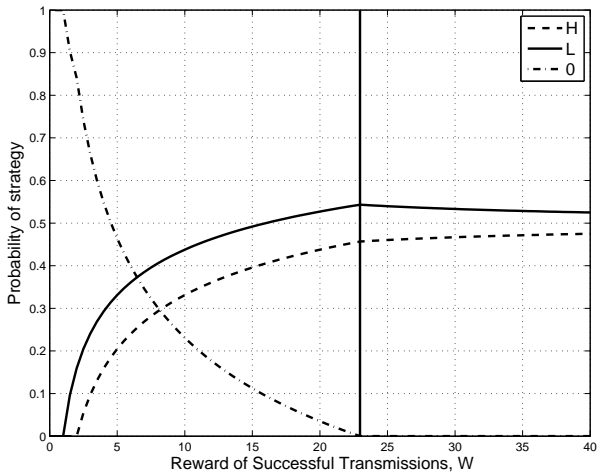}
\end{center}
\caption{The mixed strategy NE, $\sigma^* = 
(a^*, b^*, 1- a^* - b^*)$, for different values
of the reward of successful transmission, $W$, when $K = 5$.}
        \label{Fig:eplt3}
\end{figure}

Fig.~\ref{Fig:eplt3} shows the 
mixed strategy NE, $\sigma^* = 
(a^*, b^*, 1- a^* - b^*)$, for different values
of the reward of successful transmission, $W$, when $K = 5$
with $W^* = 22.969$.
Note that $W^* = 3$ when $K = 2$
according to \eqref{EQ:Ws},
and $W^*$ can be seen as the threshold value of the reward 
of successful transmission to set the probability of
$s_k = 0$ to 0. Clearly, from \eqref{EQ:Ws}, $W^*$ increases with
$K$. That is, a higher reward of successful transmission
is required to force users to keep transmitting as $K$ increases.

\begin{figure}[thb]
\begin{center}
\includegraphics[width=\figwidth]{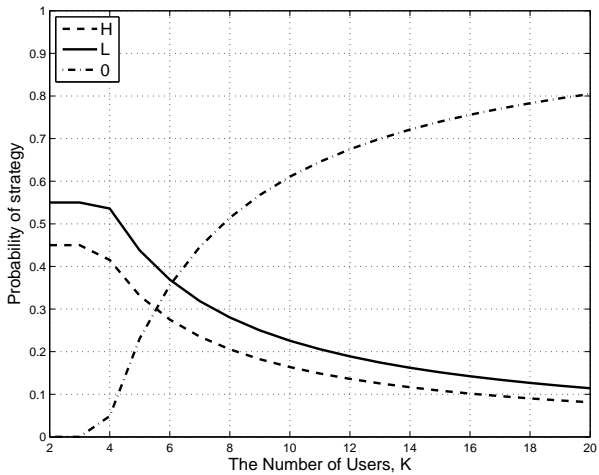}
\end{center}
\caption{The mixed strategy NE, $\sigma^* = 
(a^*, b^*, 1- a^* - b^*)$, for different numbers of users when
$W = 10$.}
        \label{Fig:eplt4}
\end{figure}

Fig.~\ref{Fig:eplt4} shows 
the mixed strategy NE, $\sigma^* = 
(a^*, b^*, 1- a^* - b^*)$, for different numbers of users when
$W = 10$. For a fixed
reward of successful 
transmissions, as the number of users increases,
the probability of transmissions (either $H$ or $L$)
decreases, while the probability of non-transmissions increases.
This behavior results from the increase
of the probability of collision as $K$ increases.

\section{Concluding Remarks}

In this paper, we formulated a multiple access game for 
ALOHA with (power-domain) NOMA,
where the payoff function is based on energy efficiency.
The mixed strategy NE has been derived using the principle of
indifference to decide 
transmission parameters, i.e.,
the probability of transmissions. 
It was shown that the probability of transmissions
can approach 1 as the reward of successful transmission
increases. In this case, unlike conventional ALOHA, we showed that
the throughput does not approach 0, although there is packet
collision because the power levels of users
can be different thanks to NOMA.

\bibliographystyle{ieeetr}
\bibliography{noma}

\end{document}